\documentstyle[12pt]{article}
\newlength{\extraspace}
\newlength{\extraspaces}
\vspace{15mm}
\pagebreak[3]
\nopagebreak
\medskip

\begin{document}

\addtolength{\baselineskip}{.8mm}

\thispagestyle{empty}

\begin{flushright}
{\sc PUPT}-1483\\
hep-ph@xxx/9407252 \\
  July  1994
\end{flushright}
\vspace{.3cm}

\begin{center}
{\Large Quantum Phenomenology for the
     Disoriented Chiral Condensate} \\

\vspace{0.4in}
{\large R. D. Amado}\\
\vspace{0.2in}
{\it  Department of Physics,  University of
 Pennsylvania \\ Philadelphia,
 PA 19104 \\
 USA} \\

\vspace{0.4in}
{\large Ian I. Kogan}
\footnote{ On  leave of absence
from ITEP,
 B.Cheremyshkinskaya 25,  Moscow, 117259,     Russia. \\
 Address after September 1, 1994: Department of Theoretical
 Physics, 1 Keble Road, Oxford, OX1 3NP, UK} \\
\vspace{0.2in}
{\it  Physics Department, Princeton  University \\
 Jadwin Hall,  Princeton, NJ 08544 \\
 USA} \\
\vspace{0.7in}
{\sc  Abstract}
\end{center}

\noindent
We consider the quantum state describing the
  Disoriented Chiral Condensate (DCC), which may be
 produced in high energy collisions.
We show how a mean field treatment of the
quantum equations corresponding to the
classical linear sigma model leads to a
squeezed state description of the pions emerging
from the DCC.  We examine various squeezed and coherent
state descriptions of those pions with particular
attention to charge and number fluctuations.
We also study the phenomenology of multiple
DCC domains.

PACS numbers:
12.38.M, 03.70

\vfill

\newpage
\section{Introduction.}

\renewcommand{\footnotesize}{\small}

\noindent

The possibility of  Disoriented Chiral Condensate (DCC) production
 in high energy hadronic or heavy ion collisions
 is a subject of much current activity.
 After the first papers
 \cite{ans}-\cite{kt} this area has attracted much interest
 in the last 2 years and  many different aspect have been discussed
 in numerous papers \cite{rw}-\cite{kogan}.
 The aim of this paper is to discuss the possible
 quantum descriptions of the disoriented chiral condensate,
 paying particular attention to its decay into pions.

     It is well known that the QCD  Lagrangian is  invariant
 (only approximately
 if nonzero masses for the light $N_{f}$  quarks are
 taken into account)
 under global chiral $SU(N_{f})_{L} \times SU(N_{f})_{R}$, where
 $N_{f}$ is the number of the light flavors. This symmetry is
 spontaneously broken to vector $SU(N_{f})_{V}$ which
 leads to $N_{f}^{2} - 1$ (quasi)goldstone bosons - pions
 (if $N_{f} = 2$) or pions, kaons and $\eta$ mesons (if $N_{f} = 3$).
 The order parameter for this breaking is the vacuum expectation
 value of  the quark condensate
 $< \bar{\psi}\psi>$. In our normal vacuum, that condensate is an
 isotopic spin zero condensate (the $\sigma$).
 However one can imagine that under some special
 conditions, for example after high-energy collision, there is
 a region in which the condensate points in some other, arbitrary
 direction in isospin space.  This
 ``cool"  region of disoriented chiral condensate, (DCC),
 would be surrounded by a ``hot"
 relatively thin expanding shell,
 which separates the internal region from ordinary  space. This picture
 was suggested by Bjorken, Kowalski and Taylor \cite{bkt}
  and is called  the "Baked Alaska" scenario.
 After hadronization the interior disoriented
 vacuum will collapse, decaying into pions. The interesting signature
 of this process could be the coherent production of either
 nearly all charged or nearly all
 neutral pions. Such remarkable charge fluctuations have been reported
  in the Centauro cosmic ray events
 \cite{centavr}. These events may be a manifestation of the DCC.
 Much of the theoretical impetus for discussions
 of the DCC have been based on classical models of QCD and of
 the classical pion field.  But the
 pions that emerge from the decay of the DCC are quanta of that
 pion field, and hence we need a quantum discussion of the field
 emerging from the DCC.  The principle results of this paper
 concern the possible forms of that quantum description.

      There are two important quantum aspects of the DCC.
The first involves the formation and evolution of the DCC and the
second its decay into pions.  The evolution of the DCC has been
discussed in terms of a simple, classical, toy model, the linear sigma
model.  We will show that a quantum treatment of that model, in the mean
field or Hartree approximation, suggest that the pions emerge from
the decay of the DCC in a squeezed state.  Other quantum states for
the pions coming from DCC decay have been proposed.
The coherent state is the first that comes to mind for quantizing
a classical field.  It has been objected that a coherent state
leads to large charge fluctuations. We will discuss various coherent
and squeezed states and show that a natural outgrowth of a real,
classical field theory o the pions is a real cartesian (in isospin)
coherent or squeezed state  and that these states do not have large
charge fluctuations.  We also discuss briefly how a few independent domains
of DCC would affect the phenomenology.

	In Section 2 we briefly review the classical linear sigma model
treatment of the DCC. This helps to establish our notation and connect
our work to the DCC literature.  In Section 3 we show how
mean field quantization of the sigma model leads to squeezed states
for the pions.  Section 4 is a discussion of the various coherent state
and squeezed state pictures that can be used to describe DCC decay.
We discuss treatments that can be found in the literature and some
new forms, with particular emphasis on solving the problem of
charge fluctuations. Section 5 briefly discusses the phenomenology
of independent domains of DCC and Section 6 presents a summary
and conclusions. A discussion of the neutral fraction probability
for classical but not spherically symmetric isospin distributions
is given in Appendix A.
Some technical material on squeezed states and
their relationship to coherent states is presented in  Appendix B.

\section{Simple Classical Model of the DCC}

Instead of considering the real hadron world with all it complexities,
most discussions of the DCC begin with a very simple, toy, classical
  model describing
 the   chiral dynamics, the linear  sigma-model.
 This expresses the dynamics in terms of a
  four component
  classical field, $ \phi^{a} = (\sigma, \vec{\pi})$,
 where   $\sigma$ and  $\vec{\pi}$ are
isoscalar and  isovector fields, respectively.  The action is
given by
\begin{eqnarray}
S = \int d^{4}x \left[ \frac{1}{2}
  \partial_{\mu}\phi^{a}\partial_{\mu}\phi^{a}
 -\frac{\lambda}{4}(\phi^{a}\phi^{a} - v^{2})^{2} + H\sigma \right]
\label{action}
\end{eqnarray}
 where $H \sim m_{q}$  describes the small explicit
  chiral symmetry breaking due to the quark masses.
 The pion mass is $m_{\pi}^{2} = H/f_{\pi} =
 \lambda(f_{\pi}^{2} - v^{2})$, where $f_{\pi} = <\sigma>$. The
 sigma meson  mass is $m_{\sigma}^{2} =  3\lambda f_{\pi}^{2}
 - \lambda v^{2} \approx 2\lambda f_{\pi}^{2}$.
 With $m_{\pi} = 135 MeV, m_{\sigma} = 600 MeV$ and
 $f_{\pi} = 92.5 MeV$ one has \cite{rw} $ \lambda = 20$ and
 $v = 87.4 MeV$.

 The fields $ \phi^{a} = (\sigma, \vec{\pi})$ parameterize the
 quark condensate
\begin{eqnarray}
 <\bar{\psi}^{i}\psi_{j}> \sim U^{i}_{j} = \sigma + i \vec{\pi}
 \vec{\tau} =
\left( \begin{array}{cc}
 \sigma + i \pi^{0} &  i\pi^{+}\\
 -i \pi^{-}  &  \sigma  - i \pi^{0} \\ \end{array}\right)
\end{eqnarray}
Let us note that if one fixes the ``absolute value" of the
 quark condensate, i.e. the determinant of the matrix
 $U^{i}_{j}$, ~$det~U = \sigma^{2} + \vec{\pi}^{2} = f_{\pi}^{2}$
 one  can get another representation, which is  used usually
  for  non-linear sigma-models:
\begin{eqnarray}
U^{i}_{j} = f_{\pi} \exp\left(\frac{i \vec{\pi}\vec{\tau}}
{f_{\pi}}\right) = f_{\pi}\cos\frac{\pi}{f_{\pi}} +
 i f_{\pi} \frac{\vec{\pi}\vec{\tau}}{\pi} \sin\frac{\pi}{f_{\pi}}
\end{eqnarray}
 where $\pi = |\vec{\pi}|$. In case of small $\pi << f_{\pi} =
 <\sigma>$ one can easily see the equivalence of these two parameterizations.

   In the usual vacuum one has $<\sigma> =
 f_{\pi}, ~ <\vec{\pi}> = 0$
 and since $\sigma$ is an isoscalar there is an  isoscalar
 condensate $<\bar{\psi}^{i}\psi_{j}> \sim \delta^{i}_{j} $
  only. However one can consider
 another configuration - $<\sigma> = f_{\pi}\cos\theta$ and
 $<\vec{\pi}> = f_{\pi} \vec{n} \sin\theta$, here  $\vec{n}$ is some
 unit vector in isospace. This condensate  corresponds
 to the DCC, i.e.
 some  classical pion field configuration
 that points in an arbitrary direction in isospin space,
 and  which is metastable.  Rajagopal and Wilczek have
 pointed out that the long wave length modes of the condensate
 grow exponentially fast after a quench \cite{rw}.
 After some time the metastable disoriented
 condensate decays  into the normal vacuum by emitting
 pions. All directions in isospin space are equally probable,
 thus there is reasonable probability that the decaying
 DCC will lead to an event with
 an unusualy large number of either neutral ($\pi^{0}$)
 or charged ($\pi^{\pm}$) pions. That would be a dramatic
 experimental signal for the DCC, a signal that may already
 have been seen in the Centauro events \cite{centavr}.
 Using the classical picture
 of DCC,  which has equal probability for all isotopic orientations,
 one finds   \cite{ans} - \cite{kt}  the probability
 $P(f) = 1/(2\sqrt{f}),~~\int_{0}^{1} P(f) df = 1$
  for the neutral pion fraction,
$f = N_{\pi^{0}}/(N_{pi^{+}}+N_{\pi^{-}}+N_{\pi^{0}})$.
Thus this picture leads to a far higher probability for
finding nearly all charged pions, $f \sim 0$ or nearly all
neutral pions $f \sim 1$ in the decay of the DCC, than
the very small probability for such distributions
that would be found
in a purely statistical decay of an excited pionic region.

The simple, classical, linear sigma model helps to motivate the
DCC and its remarkable prediction for the distribution of pion
charges in its decay. However the model is purely classical while the
decay products of the DCC are pions, the quanta of the field.  It is
therefore important to study the quantum evolution of the DCC and its
wavefunction for decay into pions.  The usual approach to quantizing
classical fields is to use coherent states \cite{Glauber}.
We will discuss the various forms of coherent states that might
be appropriate to the decay of the DCC and some of their
problems, advantages and phenomenological consequences in
Section 4, but before turning to
a description we examine models of the quantum  evolution of the DCC.
We will see in the next Section that a quantum theory of the
evolution of the DCC treated in the Hartree approximation leads
to a picture of the decaying DCC in terms of two-pion coherent
states, the so called ``squeezed states."

\section{Quantum Evolution of the DCC and Squeezed States}

As has been shown in \cite{kogan}
the most natural quantum description   of the  DCC
is given  in terms of  squeezed
states and the pion  production  in this state  is a nice
analog
of an  amplification of  oscillations in a  quantum  parametric
 oscillator.
To   get the squeezed quantum state for the  DCC  let us consider
 the mechanism for amplification of the
 long wavelength pion modes, suggested by Rajagopal and
 Wilczek in paper \cite{rw}, where the  the dynamics of
 the $O(4)$  linear sigma model after quenching was considered.
 The amplification of the long wavelength pion modes was found
in the period immediately after quenching. This  amplification
 leads to the coherent pion oscillations, i.e. to the creation
 of the  DCC. Such a behavior can be understood if one
  considers the equation of motion for the pion field \cite{rw}
 \begin{eqnarray}
 \frac{\partial^{2}}{\partial t^{2}}\vec{\pi}(\vec{k},t) +
 [\vec{k}^{2} + \lambda(<\phi^{2}>(t) - v^{2})]\vec{\pi}(\vec{k},t) = 0
  \label{rw}
  \end{eqnarray}
where we substituted (as in \cite{rw})
the $\phi^{a}\phi^{a}$  in the  nonlinear term
in (\ref{rw}) by its spatial average  $<\phi^{2}>(t)$. This is
nothing but
the Hartree or mean field approximation.
If in the initial conditions one has $<\phi^{2}>~ < v^{2}$ then the
long wavelength modes of the pion field with
$\vec{k}^{2} <  \lambda(v^{2} - <\phi^{2}>)$ start growing exponentially.
The $<\phi^{2}>(t)$  starts to oscillate near the  vacuum expectation
value $<\sigma>$ and after some time the oscillations will be damped
enough so that all the modes will be stable. Thus we
see that at the classical level, each  long wavelength mode
is described  by the equation for a parametrically excited oscillator
and one  gets the  DCC as a result of the amplification of the
zero-point quantum fluctuations of the pion field.

This picture is  similar to one which was discussed  in \cite{gs},
where
  relic graviton production  from zero-point
 quantum fluctuations during cosmological expansion was considered.
For graviton  mode with momentum $ n $ the equation
 $y''+[ n^{2} - (R''/R)] y = 0$
 was obtained, where$ R(\eta) $ is the scale factor of the
 metric $ds^{2} = R^{2}(\eta)( d\eta^{2} - d\vec{x}^{2})$ and
 a prime represents $d/d\eta$. One can see  that
this equation  is equivalent
to the pion equation (\ref{rw}) if the scale factor $R$ is connected
with $<\phi^{2}>(t)$ as $\lambda(v^{2} - <\phi^{2}>(t)) = R''/R$.

Our problem now  is to present the quantum mechanical  formulation
in  terms of
 pion creation and annihilation operators and to  get the
 wave function of the DCC.
In the  mean field   approximation the wave function
$|\Psi> = \prod_{i, \vec{k}} |\psi>_{i, \vec{k}}$ is the
product of the wave functions for each mode with momentum $\vec{k}$
and isotopic spin index $i$. Later we shall omit  $i$.
 The equation of motion (\ref{rw})
  for each mode $\pi(\vec{k},t)$  can be derived from the Lagrangian
 \begin{eqnarray}
 L_{k} = \frac{1}{2}\dot{\pi}^{2}(\vec{k},t) -
 \frac{1}{2}\Omega^{2}(\vec{k},t)\pi^{2}(\vec{k},t)
 \nonumber \\
 \Omega^{2}(\vec{k},t) = \vec{k}^{2} + \lambda
 (<\phi^{2}>(t) - v^{2})
 \end{eqnarray}
 The wave function $ |\psi>_{ \vec{k}}$ obeys the Schr\"{o}dinger equation
 \begin{eqnarray}
 i\frac{\partial}{\partial t}|\psi>_{ \vec{k}}~ =~
  H_{k}(t)|\psi>_{\vec{k}} ~ =~
 [ \frac{1}{2} {\cal P}_{\pi}^{2} +
 \frac{1}{2}\Omega^{2}(\vec{k},t)\pi^{2}(\vec{k})]|\psi>_{\vec{k}}
    \label{schr}
 \end{eqnarray}
where  $\pi(\vec{k})$ and ${\cal P}_{\pi} = -i d/d \pi(\vec{k})$
are the quantum-mechanical coordinate and momentum for the
mode with the spatial momentum $\vec{k}$.

After finding the wave function $|\psi>_{\vec{k}}$ and constructing
 the full Hartree function $|\Psi>$ one can get the value of the
 mean field $<\phi^{2}>(t) = <\Psi|\sigma^{2} + \vec{\pi}^{2}|\Psi>$
and then obtain an equation for the mean field evolution. The last
part of this selfconsistent approximation  will not be discussed in this
paper, our aim is to analyze the general features of a Hartree
wave function, taking the mean field $<\phi^{2}>(t)$ as granted.

 One can  rewrite the Hamiltonian in (\ref{schr}) in terms of
creation and annihilation operators which make it diagonal at
any given moment. It is evident that we  are interested in
getting the wave function in terms of  creation and annihilation operators
of  ordinary pions, so  we  must  diagonalize the Hamiltonian at
 $t \rightarrow \infty$, when the oscillation of the $<\phi^{2}>(t)$
 will be damped. Thus we define
\begin{eqnarray}
a(\vec{k}) = \frac{ \omega(\vec{k})
\pi (\vec{k}) + i {\cal P}_{\pi} }{\sqrt{2\omega(\vec{k})}},
{}~~~~~~
a^{\dagger}(\vec{k})  = \frac{ \omega(\vec{k})\pi(\vec{k})
- i {\cal P}_{\pi} }
 {\sqrt{2\omega(\vec{k})}}
 \end{eqnarray}
where $\omega(\vec{k})  = \Omega(\vec{k},\infty) = \sqrt{\vec{k}^{2} +
  m_{\pi}^{2}}$.
 It is easy to see that the Hamiltonian (up to a constant) is
\begin{eqnarray}
 H_{k} = \frac{1}{2}\omega(\vec{k})\bigl[1+ \frac{\Omega^{2}(\vec{k},t)}
 {\omega^{2}(\vec{k})}
 \bigr]
 a^{\dagger}(\vec{k})a(\vec{k}) -
\frac{\omega^{2}(\vec{k})-\Omega^{2}(\vec{k},t)}{4\omega(\vec{k})}
 \bigl[a^{2}(\vec{k}) + a^{\dagger 2}(\vec{k})\bigr]
 \end{eqnarray}
It is the $a^{2}(\vec{k})$ and $a^{\dagger 2}(\vec{k})$
 terms in the Hamiltonian that
transform the initial vacuum  $|0>$  into a
 squeezed state $S(r,\phi)|0>$. The squeezed state is given by  (\cite{yuen})
\begin{equation}
|\xi> = S(\xi)|0> =
\exp\left[\frac{1}{2}\left(\bar{\xi}a^{2} - \xi (a^{\dagger})^2)
\right)\right]|0>
 \label{s}
 \end{equation}
and we are writing  $\xi = r \exp(i\phi)$. More generally, one can
consider the squeezed coherent state  $ S(\xi) \exp (\alpha a^{\dagger})|0>$.
Some technical details
 about squeezed states are presented in Appendix B.

 To calculate the squeezing and phase  parameters $r$ and $\phi$  let us
  consider the solution of the Schr\"{o}dinger equation
(\ref{schr}) in the coordinate (here it is $\pi$) representation
( we omit the mode label $\vec{k}$ for a  moment)
\begin{eqnarray}
 <\pi|\psi>(t) = C(t) \exp (- B(t) \pi^{2} + D(t) \pi)
\label{wavefunction}
\end{eqnarray}
 where we also take into account the possiblity of the  time-dependent
 expectation value of the $\pi$ field   $<\pi(t)> =
 <\psi(t)|\pi|\psi(t)> \sim D(t)/2B(t)$. This state corresponds to the
squeezed  coherent state.
For $B = \omega/2$ this wave function describes the  usual
 coherent state and, if $D = 0$, it is the vacuum
  state. For all other values this wave function describes
 the squeezed state (\ref{s}) where the parameters $r$ and $\phi$
 are connected to $B$   by the relation \cite{yuen} (see
 also  \cite{sqrev1} - \cite{gs})
 \begin{eqnarray}
 B = \frac{\omega}{2} \frac{\cosh r + \exp(2i\phi)\sinh r}{
 \cosh r - \exp(2i\phi)\sinh r} \nonumber \\
\cosh 2r = \frac{ \omega^{2} + 4 |B|^{2}}{4\omega ReB}; ~~~
\sin 2\phi = \frac{1}{\sinh 2r} \frac{Im B}{ Re B}
\label{B}
\end{eqnarray}
 Substituting (\ref{wavefunction}) into (\ref{schr})  one gets
  equations for $B(t)$, $C(t)$ and $D(t)$:
  \begin{eqnarray}
  i\dot{B}(t) = 2B^{2}(t) - \frac{1}{2}\Omega^{2}(t), ~~~~~
  i\dot{D}(t) = 2B(t)D(t),~~~~~
  i \frac{\dot{C}(t)}{C(t)} = B(t) - \frac{1}{2} D^{2}(t)
  \end{eqnarray}
  which means that $B(t)$ is related to the solution
   of the classical equation (\ref{rw})
   \begin{eqnarray}
  B(t) = - \frac{i}{2} \frac{\dot{\psi}(t)}{\psi(t)}, ~~~~~~~~~~~~~~
   \ddot{\psi}(t) + \Omega^{2}(t) \psi(t)  = 0
   \label{Bpsi}
   \end{eqnarray}
 in terms of which $D(t)$ can also be obtained
  \begin{eqnarray}
D(t) = D(0) \exp(-2i\int_{0}^{t}B(\tau)d\tau) =
D(0) \exp\left( -\int_{0}^{t}\frac{\dot{\psi}(\tau)}{\psi(\tau)}
d\tau\right) = D(0) \frac{\psi(0)}{\psi(t)}
\label{D(t)}
\end{eqnarray}
and $C(t) = C(0)\exp\left(i\int_{0}^{t}(B(\tau)-D^{2}(\tau)/2)d\tau\right)$
 is an evident  phase factor.
 The  equation (\ref{Bpsi})  can be viewed as a Schr\"{o}dinger equation
 describing the wave function $\psi(t)$  of  a  ``particle"
 with mass $m = 1/2$
 on a line $t$ having energy $\omega^{2}(k) = k^{2}+m_{\pi}^{2}$
and moving through the
potential barrier $V(t) = - \lambda
(<\phi^{2}>(t) - f_{\pi}^{2})$
 \begin{eqnarray}
-\frac{d^{2}\psi(t)}{dt^{2}} + \lambda (f_{\pi}^{2} - <\phi^{2}>(t))
\psi(t) =
 \omega^{2}(k) \psi(t)
\end{eqnarray}
 Far from the barrier, i.e. at $t \rightarrow \pm \infty$ one
has $V(\pm \infty) = 0$ and the general solution
 of the Schr\"{o}dinger equation at $ t \rightarrow \pm \infty$
 is a superposition of  left  and right moving waves
 \begin{eqnarray}
\psi^{+}(t) = S^{+}_{R} e^{-i\omega(k) t}  + S^{+}_{L} e^{+i\omega(k) t},
{}~~~~~~~~ t \rightarrow +\infty
 \nonumber \\
 \psi(t) = S^{-}_{R} e^{-i\omega(k) t}  + S^{-}_{L} e^{+i\omega(k) t},
 ~~~~~~~~ t \rightarrow  -\infty
 \label{abcd}
 \end{eqnarray}
 Due to  unitarity the  total fluxes at $t \rightarrow \pm \infty$
 must be equal  $|S^{+}_{L}|^{2}-|S^{+}_{R}|^{2}=
|S^{-}_{L}|^{2}-|S^{-}_{R}|^{2}$ and one
 can find
\begin{eqnarray}
 S^{+}_{R} & = &\cosh r S^{-}_{R} - e^{2i\theta} \sinh r S^{-}_{L}
 \nonumber \\
 S^{+}_{L} &  = & - e^{-2i\theta} \sinh r S^{-}_{R} +   \cosh r S^{-}_{L}
 \label{ABCD}
 \end{eqnarray}
 where $\theta$ is the scattering phase and the factor $r$
 is defined by the  probability of transition through
 the barrier.

 Let us remember that if we are starting
  from the vacuum at $t \rightarrow -\infty$, i.e. from $B = \omega/2$,
 one must have $S^{-}_{R} = 0$. Even if we assume that the initial
 state is not the vacuum, but some coherent state with nonzero $D(0)$
 we still have  to put  $B = \omega/2$ initially.
This means that  at the left
(large negative $t$) we have only a left moving outgoing wave
 $S^{-}_{L} e^{+i\omega(k) t}$.  At the
 right (large positive $t$) one has both left and right moving waves,
  i.e.
 the incoming  $S^{+}_{L} e^{+i\omega(k) t}$
 and reflected $S^{+}_{R} e^{-i\omega(k) t}$ waves.  The transition
 coefficient can be obtained from  (\ref{ABCD}) by puting $S^{-}_{R}=0$
 \begin{eqnarray}
  \frac{|S^{-}_{L}|^{2}}{|S^{+}_{L}|^{2}} = \frac{1}{\cosh^{2} r}
 \end{eqnarray}

  Now let us calculate
$B(t) = -(i/2) (\dot{\psi}/\psi)$
 at large positive $t$.  Using (\ref{abcd})  one  can find
 after simple calculation (restoring the $k$ dependence) :
 \begin{eqnarray}
 B(k) = \frac{\omega(k)}{2} \frac{\cosh r(k) + \exp[2i(\theta -\omega(k) t)]
 \sinh r(k)}{\cosh r(k) - \exp[2i(\theta - \omega(k) t)]\sinh r(k)}
 \end{eqnarray}
 in complete agreement  with (\ref{B}), with  the phase factor
 $\phi(t)$   depending on time as
$\phi = \theta - \omega(k)t$.

The squeezing parameter $ r(k)$ depends on the absolute
 value, $k$,  of the  mode spatial  momentum $\vec{k}$
 and    is determined by the  probability
 of tunneling through the  potential barrier
 $V(t)=\lambda\bigl[f_{\pi}^{2} - <\phi^{2}>(t)]$.
 Let us note that tunneling takes place precisely when
 $\omega^{2}(k)  - V(t) <0$, i.e. when the classical long
 wavelength modes  are exponentially amplified
and we see once more that  squeezing is ultimately
 connected with the exponential growth of the
 classical long wavelength modes
 and the squeezing for each mode $k$  is determined by the
 function $<\phi^{2}>(t)$. This is the only input information
 we must know to calculate the  DCC  wave function.

After calculating $B(t)$ one can study the effect of a displacement $D(t)$.
 All we need to know is the ratio $\psi(t)/\psi(0)$, where ``$0$" is an
 arbitrary initial time and we shall put it to
 $t_{-}\rightarrow-\infty$.
 It is easy to see that (we put the scattering phase $\theta = 0$)
 \begin{eqnarray}
 D(t_{+} \rightarrow \infty) =
 D(t_{-} \rightarrow -\infty)\frac{S^{-}_{L} \exp(+i\omega(k) t_{-})}
 { S^{+}_{R} \exp(-i\omega(k) t_{+})  + S^{+}_{L} \exp(+i\omega(k) t_{+})} =
 \nonumber \\
 e^{-i\omega(k) (t_{+}-t_{-})}\frac{1}{\cosh r
 \left(1 -  \exp(-2i\omega(k) t_{+})\tanh r\right)}
 \end{eqnarray}
  and one can see  that at large $t_{+}$, on average,  $D(t_{+})$ is small
 due to the suppression factor $1/\cosh r$. Only at moments
 $t = \pi n/\omega$  during the short intervals
 $\delta t \sim  e^{-2r}/\omega$ does one get large $D(t)$. However
 these pulses may be an artefact of the approximation,
 with the average
  $D(t)$ being of order $e^{-r}$. Thus we see that for those
modes for which  the squeezing is large, the initial displacement, i.e.
the coherent part of the wave function, disappears at late times
 independently of how large it was in the beginning.
 The  mean value of the pion field $<\pi(t)> \sim D(t)/B(t)$
 and one gets
\begin{eqnarray}
 <\pi(t)> \sim
 e^{-i\omega(k) (t_{+}-t_{-})}\frac{1}{\cosh r
 \left(1+ \exp(-2i\omega(k) t_{+})\tanh r\right)}
 \end{eqnarray}
  and  on average,  $ <\pi(t)>$ is  also suppressed
  by a  factor $1/\cosh r$. Only at moments
 $t = (\pi n + \pi/2)/\omega$  during the short intervals
 $\delta t \sim  e^{-2r}/\omega$ does one get large $ <\pi(t)>$. However
 these pulses may be also  an artefact of the approximation we use
 as the
 pulses for $D(t_{+})$ and on average the classical pion field is small.
 This is a very important
  point demonstrating  that for  low momentum pion modes, the classical
 pion field, even if it were produced at some intermediate step, will
 ultimately be
 negligible.
  This is for the displacement field.  The average of the
square of the pion field, $< (\pi(t))^2 >$ is, of course, not
small but is the exponentially growing condensate of the DCC.
This is the part that leads to the squeezed state.

 In the quasiclassical
 approximation it is easy to calculate the squeezing parameter
  $r(k)$:
  \begin{eqnarray}
  r(k) = 2 Re  \int dt \sqrt{\lambda
  \bigl[v^{2} - <\phi^{2}>(t)\bigr]  - k^{2}}
  \end{eqnarray}
  which is valid for  small $k$  when  $r(k)>>1$
  The average number of particles in each mode is
$
<N_{k}> = \sinh^{2} r(k)$
and we see that $<N_{k}>$ sharply decreases with the increase of $k$.
 To make a rough estimate
  let us consider a simple model for $<\phi^{2}>(t)$. We assume
   that $<\phi^{2}>(t) = 0$  in the interval $t \in (0, \tau)$ and
is  equal to its usual v.e.v. $<\phi^{2}>(t) = f_{\pi}^{2}$ outside
  this interval. Such  behavior for $<\phi^{2}>(t)$
  is a very rough approximation to a more realistic behavior obtained
in \cite{rw} in numerical experiments, nevertheless one can use
it for some preliminary estimates.   The transmission coefficient in this case
 is well known (see for example \cite{LL}) and in our notation
 takes the form
 \begin{eqnarray}
  \frac{1}{\cosh^{2} r(k)} =
 \frac{4\omega^{2}(k)(m_{\sigma}^{2}/2 - \omega^{2}(k))}
 {4\omega^{2}(k)(m_{\sigma}^{2}/2 - \omega^{2}(k)) +
 (m_{\sigma}^{2}/2)^{2} \sinh^{2} \tau \sqrt{m_{\sigma}^{2}/2
 - \omega^{2}(k)}}
 \end{eqnarray}
 where we used the relation $m_{\sigma}^{2} = 2\lambda f_{\pi}^{2}$.
 Then the average number of particles with momentum $k$  is
 \begin{eqnarray}
  <N (k)> = \sinh^{2} r(k) =
  \left(\frac{m_{\sigma}^{2}}{2}\right)^{2}
  \frac{\sinh^{2} \tau \sqrt{m_{\sigma}^{2}/2
 - \omega^{2}(k)}}{4\omega^{2}(k)(m_{\sigma}^{2}/2 - \omega^{2}(k))}
 \label{N(k)}
   \end{eqnarray}

  One can estimate the average number of particles  $<N (k)>$ for different
  $k$. For $\tau$ we shall use an estimate $\tau = 3-6 ~m_{\sigma}^{-1}$
which comes from the results of \cite{rw} $\tau = 1-2 ~ (200 MeV)^{-1}$.
 Then for $k =0$ one gets  $<N(0)> \approx 10^{2} - 10^{3}$, at
 $k = m_{\pi}$  the $<N(m_{\pi})>$ is approximately twice as small
 but at $k = 3m_{\pi} \approx 400 MeV$ it is  two orders of magnitude
 smaller $<N(3m_{\pi})>\approx 1 - 10$.
 Of course this is the very crude estimate, however it demonstrates
 the qualitative features of the phenomenon - the
  sharp exponential dependence on $k$ and large amplification
factor in $r$ which is of order of $m_{\sigma}\tau$, where
$\tau$ is the characteristic time of damping of
$<\phi^{2}>(t)$ oscillations.

In modeling a realistic ``Baked Alaska" scenario one should
 include the effect of the expansion. In \cite{rw} it was suggested
  that the expansion be described by including
the term $\dot{a}\pi/a$ in the equations
 of motion, where $a(t)$ is a scale factor for the expanding plasma.
 This is the same as
considering the problem in a space-time
with metric $ds^{2} = dt^{2} - a^{2}(t)d\vec{x}^{2}$. It is
easy to show that choosing the new time  $\tilde{t}(t) = \int dt/a^{3}(t)$
one gets the same  Schr\"{o}dinger equation as in (\ref{schr}) but with
new $\tilde{\Omega}^{2}(k,\tilde{t}) = a^{6}(\tilde{t})\Omega^{2}
(k,\tilde{t})$. Thus taking into account the expansion one gets
squeezed state again but with parameters which depend on the  details
of an expansion - the scale factor $a(t)$.
In a case of
anisotropic  $1+1$ expansion,
which seems more  realistic at early stages of
the DCC formation
(if any)  and which was considered  in
\cite{khlebnikov}, \cite{kogan1}  and later in \cite{lanl}
  the  convenient
 coordinates parameterizing the interval are proper time $\tau$
and rapidity $\eta$
\begin{equation}
\tau = \sqrt{t^{2}-x^{2}},~~~\eta =
  \frac{1}{2}\ln\left(\frac{t-x}{t+x}\right)
  \end{equation}
  and $ds^{2} = d\tau^{2} - \tau^{2}d\eta^{2} - dx_{\perp}^{2}$
  In this case one gets the Schr\"{o}dinger equation  for evolution
 in $\tau$ with the ${\Omega}^{2}(k_{\perp}, k_{\eta},
 \tau)$ where momentum $k_{\eta}$   appears in the  combination
 $k_{\eta}^{2}/\tau^{2}$, which means that at small $\tau$
one has amplification only at  very low $k_{\eta}$.  One
 can get  analogs of (\ref{N(k)}). In this case
  it will be $N(k_{\eta}, k_{\perp})$ and one can
  get the distribution in rapidities $\eta$ after taking the
  Fourier transform in $k_{\eta}$.

Thus we see that quantizing the linear sigma model for the DCC and
treating the non-linear equation in the Hartree or mean field
approximation leads to a simple picture of the quantum DCC that
suggest a squeezed state treatment of the
emerging pion waves. We also find the exponential amplification
of the long wavelength modes that is seen classically leading, in
a simple model, to a very large number of low energy
coherent pions. We now turn to various treatments for the wave
function of those emerging pions.

\section{Quantum Wavefunctions for Decay of the DCC}

There are many ways to go from a classical pion field theory
to a quantum state, and many of these have appeared in the
literature of DCC and other places.  In this Section, rather that only give
the squeezed state description we developed in Section 3,
we present and contrast  these various treatments.
Thus this section does not depend on the squeezes state scenario, but
rather discusses a range of possible quantum treatments of the classical
decaying DCC.

The simplest quantum description of a classical field is given in terms
  of a coherent state \cite{Glauber}
\begin{eqnarray}
|{\bf \pi}> = \exp\left(\frac{-{\bar N}}{2}
\right) \exp\left(\int d^{3}\vec{k}{\bf{\pi}}(\vec{k})\cdot
{\bf a}^{\dagger}(\vec{k})\right)|0>
\label{chst}
\end{eqnarray}
 where $ {\bf {\pi}}$ is the isotriplet pion field and
 and ${\bf a}^{\dagger}$ are corresponding creation operators.
 The mean number of pions in the state is given by
 \begin{equation}
{\bar N}= \int d^{3}\vec{k}{\bf{\pi}}\cdot {\bf{\pi}}^{*}
\end{equation}
which number also comes into the normalization of the coherent state.
The root mean square number fluctuation in this state is $\sqrt{{\bar N}}$,
and hence the fractional number fluctuation decreases for large ${\bar N}$.
If one defines position and momentum like variables
for each mode, $\vec{k}$, by
\begin{equation}
{ \vec {r}}  =  \frac{1}{\sqrt{2}} ({\bf a}^{\dagger}+ {\bf a})
\label{are}
\end{equation}
and
\begin{equation}
 { \vec {p}}  =  -i \frac{1}{\sqrt{2}} ({\bf a}^{\dagger}- {\bf a})
 \label{pee}
\end{equation}
one also finds that the coherent state is a minimal uncertainty state.
It is these two properties, minimal uncertainty and small number fluctuation
in the large ${\bar N}$ limit, that make the coherent state attractive as
the quantum representative of the classical field.

  However this coherent state wave function has been criticized  in
 \cite{kt} because, for arbitrary ${\bf{\pi}}(\vec{k})$
 such a description leads to
 large charge  fluctuations in the pion system.
 In the remainder of this section we will discuss
 ways of addressing this criticism while maintaining as much
 as possible of the advantages of the coherent state, minimum uncertainty
 and small number fluctuations, and also maintaining the relatively large
 probability of finding a neutral pion fraction near zero or one.
 We will divide our discussion in two parts, states with identically
 zero charge and charge fluctuations, and states with only average
 charge zero.

	Let us begin with states with average charge zero but still
having charge fluctuations.  Consider the coherent state defined
in (\ref{chst}).  Let us use a cartesian rather than a spherical tensor
representation for the isospin of the
pions, both for the creation and annihilation
operators and for the classical field, ${\bf{\pi}}(\vec{k})$.
Since the classical field equations are real, we might expect that
classical field configuration to be real as well.
Hence a real cartesian quantum description is a natural
outgrowth of the classical beginnings of the DCC. We will now show that
this leads to zero expected charge in the coherent state and small
fractional charge fluctuations.  In terms of cartesian pions, the
number operator for $\pi^+$ is given by
\begin{equation}
n_+ = \frac{1}{2}(a_x^{\dagger} a_x + a_y^{\dagger} a_y) +\frac{i}{2}
(a_y^{\dagger} a_x - a_x^{\dagger} a_y)
\end{equation}
and for $\pi^-$ by
\begin{equation}
n_- = \frac{1}{2}(a_x^{\dagger} a_x + a_y^{\dagger} a_y) -\frac{i}{2}
(a_y^{\dagger} a_x - a_x^{\dagger} a_y)
\end{equation}
so that the charge, $Q=n_+ - n_-$ is
\begin{equation}
Q = i( a_y^{\dagger} a_x - a_x^{\dagger} a_y)
\end{equation}
This is for each mode, ${\vec {k}}$, the total charge is the
integral over ${\vec {k}}$. It is easy to see that, for real cartesian
pion field, the expected value of the charge in the coherent state
is zero.  The expected value of the charge squared is
\begin{equation}
 < Q^2 >= \int  d^{3}\vec{k} ( {\bf \pi}^2_x + {\bf \pi}^2_y )
 \end{equation}
 which is a number of order ${\bar N}$.  Hence the fractional root mean
 square charge fluctuations are of order $1/\sqrt{{\bar N}}$ and therefore
 small in the large ${\bar N}$ limit. It is clear that the real cartesian
 coherent state also continues to be a minimal uncertainty state.

	We would expect that in the DCC the classical pion field will
point in some fixed direction in isospin space for all  ${\vec {k}}$.
Suppose we label that direction with the usual polar angles, $\theta$,
$\phi$, and continue to use the real cartesian coherent state.
Then it is easy to see that the average number of $\pi^0$'s
is given by ${\bar N} \cos^2 \theta$.
The neutral pion
fraction is thus given by $f = \cos^2 \theta $, and corresponds to
a probability of $P(f) = 1/(2\sqrt{f})$, as before.   The root mean square pion
charge fluctuation is given by $\sqrt{{\bar N}} \sin \theta$ and the
fractional charge fluctuation  by $ \sin \theta/\sqrt{{\bar N}}$,
which is small
in the large ${\bar N}$ limit.  Hence the real cartesian coherent state
has many desirable features for describing the radiation of the DCC.  Its
major faults are that it does not have identically zero charge and that
it is not well linked to a simple quantum theory of the DCC dynamics itself,
which suggests squeezed states. \cite{kogan}  We now turn to these issues.

	Let us begin with the application of squeezed states to the
problem of decay of the DCC. We first review some features of
squeezed states.
For a single mode created by the operator
$b^{\dagger}$, the normalized squeezed state may be written
\begin{equation}
|\alpha> = (1-|\alpha|^2)^{1/4} \exp \left(
\frac{\alpha}{2} (b^{\dagger})^2 \right) |0>
\end{equation}
where $\alpha$ is a complex number with modulus less than one
and is related to the parameters of (\ref{s}).
 (See Appendix B for more details.)
The squeezed state is a minimum uncertainty state, for position and
momenta defined in terms of the operators $b$ as in (\ref{are})
and (\ref{pee}).
In the case of the coherent state, that minimum uncertainty is achieved
by having each of $\Delta x$ and $\Delta p$ take on their minimum values.
For the squeezed state, only their product need be minimum, and one can
be large while the other is ``squeezed" to keep the product fixed. An
important difference between the squeezed state and the coherent state
comes in the number fluctuations.  For the squeezed state we find
\begin{equation}
<\alpha| (b^{\dagger} b)^2 -({\bar N})^2|\alpha > =
2 {\bar N} ({\bar N} +1)
\end{equation}
which means that the fractional root mean square fluctuations are of
order 1, in the large ${\bar N}$ limit rather than of order
$1/\sqrt{ \bar N}$ as they are in the coherent state case.

	In spite of the large number fluctuations in the squeezed state,
let us look at its application to the DCC. Suppose we construct the
operator $b^{\dagger}$ from a real isospin rotation of the cartesian
pion operators
\begin{equation}
b^{\dagger} = \cos \theta a^{\dagger}_z + \sin \theta \cos \phi a^{\dagger}_x
+  \sin \theta \sin \phi a^{\dagger}_y
\end{equation}
where $\theta$ and $\phi$ are the usual polar angles.  We may now
label our squeezed state $| \alpha, \theta,\phi>$.  It is clear that
the mean number of pions in the state is still ${\bar N} =
\frac{|\alpha|^2}{1-|\alpha|^2}$. The expectation of
the total charge in this state
is easily seen to be zero.  The expected value of $\pi^0$'s is given by
\begin{equation}
<\alpha , \theta,\phi| a^{\dagger}_z a_z|\alpha , \theta,\phi> = {\bar N}
 (\cos \theta)^2
 \end{equation}
 The neutral pion fraction is $f = \cos^2 \theta$ and its probability
 is again $P(f) = \frac{1}{2\sqrt{f}}$.  Hence the squeezed state with
 arbitrary real cartesian isospin direction has zero average charge,
 minimum uncertainty, and a large probability for neutral fraction near
 zero or near one.  However it has large (of order ${\bar N}$) charge
 fluctuations.

	To deal with the charge fluctuations, it was
 suggested in \cite{kt} that the  quantum state
 that describes the DCC be an isosinglet. A particular isospin zero
 wave function was considered  a long time ago
 by Horn and Silver \cite{hs}.
Consider the operators
 $A$ and $A^{\dagger}$, with
\begin{eqnarray}
A^{\dagger} = 2 a_{+}^{\dagger}a_{-}^{\dagger} - (a_{0}^{\dagger})^{2}
=-((a_x^{\dagger})^2 + (a_y^{\dagger})^2 +  (a_z^{\dagger})^2)
\end{eqnarray}
which is clearly an isoscalar operator.  Any function of $A^{\dagger}$
operating on the vacuum will create a state with isospin zero and thus
an eigenstate of charge with zero eigenvalue.  Such a state will
have not just zero average charge, but zero charge fluctuations as well.
We note, in passing,  that  $A, ~ A^{\dagger}$ and $N =
\sum_{i}a_{i}^{\dagger}a_{i}$
  are the generators of an $sl(2)$ algebra
  \begin{eqnarray}
  [A, A^{\dagger}] = 4N + 6,~~[N,A] = - 2A,~~[N,A^{\dagger}] = 2A^{\dagger}
  \end{eqnarray}

	A particularly simple state of isospin zero and exactly
$2N$ pions has been suggested by a number of authors, cf. \cite{kt}.
It is of the form
\begin{eqnarray}
|\Psi> = \frac{1}{\sqrt{(2N+1)!}} (A^{\dagger})^N |0>
\end{eqnarray}
Using Stirling's formula for the asymptotics of $n! \sim \sqrt{2n}
\exp(n\ln(n/e))$ it is easy to see
  that  the probability of having 2n neutral pions
  in the state $|\Psi>$ is \cite{hs}, \cite{kt}
\begin{eqnarray}
P(n,N) = \frac{(N!)^{2} 2^{2N}}{(2N+1)!}\frac{(2n)!}{(n!2^{n})^{2}}
 \sim \sqrt{N/n},~~n,N >>1
\end{eqnarray}
and corresponds to the $1/\sqrt{f}$ distribution of
the classical picture.

The same distribution would be obtained
if an  arbitrary  relative phase factor were
inserted between the charged
 and neutral creation operators
    $\left(2 a_{+}^{\dagger}a_{-}^{\dagger} -
  \exp(i\theta)(a_{0}^{\dagger})^{2}\right)^{N}$
This might seem to imply that the  zero isospin
 condition  is not
 so important, and that one can
 introduce any arbitrary factor.
 This is not the case.  With any arbitrary factor in front of
$(a_{0}^{\dagger})^{2}$, the states created will still have
the third component of isospin equal to zero and hence be
zero eigenstates of charge, but the operator $A^{\dagger}$ will
now be a combination of isoscalar and second rank iso-tensor
All directions in iso-space will not be equivalent and we
will no longer have a $1/\sqrt{f}$ neutral fraction distribution.
For example if the arbitrary factor is zero, there will be no neutral fraction.

	The state $|\Psi>$ considered above is quite restrictive.
It has exactly $2N$ pions. A quantum state arising from a classical
field is not expected to have a sharp number of quanta.
We now consider two isoscalar states
that are closer in spirit to the coherent state, and have an
indefinite number of pions.  First consider the squeezed state
made by exponentiating $A^{\dagger}$.
We write
\begin{equation}
|\Psi,\alpha > = (1-|\alpha|^2)^{\frac{3}{4}} \exp \left(\frac{\alpha}{2}
A^{\dagger}\right) |0>
\end{equation}
To understand the normalization factor note that this state can be
written in terms of cartesian pions as a product of three simple
squeezed states,
\begin{equation}
|\Psi,\alpha > = (1-|\alpha|^2)^{\frac{3}{4}}
\exp\left( \frac{-\alpha}{2}((a_x^{\dagger})^2
+(a_y^{\dagger})^2 + (a_z^{\dagger})^2)\right)|0>
\end{equation}
The probability of finding  $2n$ neutral pions in this state is
\begin{eqnarray}
P(n) =  \sum_{N = 0}^{\infty}\frac{1}{(N!)^{2}} P(n,N) \sim
 \frac{1}{\sqrt{n}}
 \end{eqnarray}
 - again the same distribution. Although this state is a minimum uncertainty
 state, and
 there are no charge fluctuations
 in it, there are large particle number fluctuations.
The number operator can be written in terms of the cartesian
number operators as $n_x+n_y+n_z$.  Its expectation value in
the state $|\Psi, \alpha>$ we call ${\bar N}$.
Using the product nature of the
state we can then calculate
\begin{equation}
\sigma^2 = <\Psi, \alpha|(n_x+n_y+n_z)^2|\Psi, \alpha> - {\bar N}^2
\end{equation}
using the results for a simple squeezed state.  We find
\begin{equation}
\sigma^2 = \frac{{\bar N}^2}{3} +2{\bar N}
\end{equation}
This shows that the number fluctuations grow like ${\bar N}$, or the
fractional fluctuations like $1$ rather than like $1/\sqrt{{\bar N}}$.

	As an alternative for the iso-scalar state one can consider
the iso-spin projected coherent state first introduced by Botke, Scalapino
and Sugar, \cite{BSS} and recently used to describe the coherent pions
emerging from nucleon-antinucleon annihilation. \cite{Amado1}, \cite{Amado2}
The $I = 0$ projected  coherent state is
\begin{equation}
| \lambda ,0>= {\cal {N}} \int d {\hat T} e^{\lambda(k) d^3k
a^{\dagger}_k \cdot {\hat T}}
|0> Y^{\star}_{0,0} ({\hat T})
\end{equation}
where the normalization is given by
\begin{equation}
{\cal {N}}= (4 \pi j_0 (-i{\bar N}))^{-1/2}
\end{equation}
and the mean number of particles, ${\bar N}$ is given by
\begin{equation}
{\bar N}= \int d^3k \lambda (k) \lambda^{*} (k)
 \end{equation}
 This is an iso-scalar state and hence has zero average charge and zero
charge fluctuation.  It is also easy to show that for large ${\bar N}$,
the variance, $\sigma$ grows like $\sqrt{{\bar N}}$ so that the fractional
particle number fluctuations decrease for large ${\bar N}$.
However the state $| \lambda, 0>$, is not a minimum uncertainty
state.
To study the uncertainty in the state
let us look at the uncertainty in the third component of
isospin (all components are equivalent since we are in an $I=0$
state). Define
\begin{equation}
x_0 = \frac{1}{\sqrt{2}}(a_0 + a_0^{\dagger})
\end{equation}
and
\begin{equation}
p_0 =-i \frac{1}{\sqrt{2}}(a_0 - a_0^{\dagger})
\end{equation}
where the subscript zero refers to the third component of isospin.
It is easy to see that
\begin{equation}
<\lambda,0|x_0|\lambda,0> = <\lambda,0|p_0|\lambda,0>=0
\end{equation}
This is also a feature of the $I=0$ squeezed state.
We need to look at $x^2$ and $p^2$.  They are given by
\begin{equation}
x_0^2 = \frac{1}{2}( a_0^2 + (a_0^{\dagger})^2) +a_0^{\dagger} a_0 +1/2
\end{equation}
and
\begin{equation}
p_0^2 = -\frac{1}{2}( a_0^2 + (a_0^{\dagger})^2) +a_0^{\dagger} a_0 +1/2
\end{equation}
One finds (taking $\lambda(k)$ as real)
\begin{equation}
<\lambda,0|a_0^2|\lambda,0> =<\lambda,0|(a_0^{\dagger})^2|\lambda,0> =
\frac{{\bar N}}{3}
\end{equation}
and
\begin{equation}
<\lambda,0|a_0^{\dagger} a_0|\lambda,0> = \frac{{\bar N} }{3}
\frac{ i j_1(-i{\bar N})}{ j_0(-i{\bar N})}
\end{equation}
Finally in the large ${\bar N}$ limit one obtains
\begin{equation}
\Delta x \Delta p =  \frac{{\bar N}}{3}
\end{equation}
Hence  the isospin projected coherent state is
not a minimal uncertainty state.

Thus we see that there are many classes
  of trial wave functions that can reproduce some of the
 phenomenologically desirable features of DCC decay.  It is natural to ask
 the question what is  the most natural
 class of these functions and what are the dynamical  mechanisms
 leading to the   generation of these functions.
In the previous section we argued that simple quantum
dynamics for the evolution of the DCC suggests the squeezed state
as being most natural.
In this section we have presented a number of quantum state
candidates for discussing the pion state from the decay of the DCC.
We have examined them from the point of view of charge fluctuations,
number fluctuations, minimal uncertainty and neutral fraction probability.
None of the candidates pass all criteria, but the cartesian
pion squeezed
state and the cartesian pion coherent state seem the most attractive.
The first has the better dynamic credentials, but it also has much larger
number fluctuations than the second.

\section{Domains of DCC}

	We have seen that most reasonable quantum treatments of the
pions from decay of the DCC lead to a probability $P(f)$ of
neutral pion fraction $f$ of $P(f) =1/(2\sqrt{f})$. This means that
the probability of finding nearly all charged pions or nearly all
neutral pions is quite large compared with what it would be for a
purely random collection of $N$ ($N$ large) pions.  The probability
of finding $f$ within $\epsilon$ of zero (corresponding
to nearly all charged pions)
is $\sqrt{\epsilon}$,
while the probability of finding
$f$ within $\epsilon$ of one (nearly all neutral pions) is $\frac{1}{2}
\epsilon$ for small $\epsilon$.   It is these surprisingly large probabilities
that has fueled much of the interest in DCC, since they imply a strong
and unambiguous
experimental signal for the DCC. But in reactions in which a DCC can be
created it is possible that not a single domain of DCC will be created but
that perhaps a number of disconnected domains will be created each with
a separate random direction in isospin space for the condensate.
What then will be the probability of an overall neutral pion
fraction $f$?

	Suppose there are
a number of regions, domains, or bubbles of DCC and that the
isospin condensate direction in
 each is arbitrary.  If, of course, there are many many such domains
 the average of $f$ over the domains will be $1/3$ with very small
 variance.  In particular the chance of finding $f$ near zero would
 be very very small.  But  a more likely
 scenario is that there may be a few domains, more perhaps than one, but not
 many many.  Then the probability of finding small $f$ is not
 prohibitively small.
 For example, suppose there are two domains, and we average the
 pion observations over both.
 Then,  the probability of finding neutral fraction $f$ is
 $\pi/2$ for $f <1/2$ and $\pi/2 -2 \arccos(\frac{1}{\sqrt{2 f}}$
 for $1/2<f<1$. Similar forms can be easily obtained for $n$ domains.
 In particular one sees that the probability of finding neutral fraction
 less that $\epsilon$ is $\sqrt{\epsilon}$ for the one domain case, and
 $(\pi/2) \epsilon$ in the two domain case.  For $n$ domains the
 probability of finding neutral fraction less than $\epsilon$ goes
 like $\epsilon ^{\frac{n}{2}}$.  For reasonably small numbers of
 domains this is not ridiculously small.

Hence one might build a phenomenological model of how the
DCC manifests itself
in terms of a {\bf small} number of statistically
independent domains of DCC.  That might be closer to what happens
in heavy ion collisions, say, and yet  holds out hope of showing
an interesting signal.

\section{Summary and Conclusions}

The idea that a Disordered Chiral Condensate (DCC) might occur
in a rapid quench after a high energy collision is an attrative one.
The signature of a DCC would be the emission of a large number of pions
either nearly all neutral or nearly all charged.  Such a remarkable
process may have already been seen in the Centauro events.\cite{centavr}
Most discussions of the DCC concentrate on how it arises in a simple
classical field theory, the linear sigma model, and how the long
wave length modes grow until they encounter the hot boundary of
the quenched region (``Baked Alaska").
But the particles detected from the DCC are pions, the quanta of the
field, and hence a quantum discussion both of the growth of the
DCC and of its deacy is needed.  In this paper we have attempted
to provide the beginnings of such a discussion.
By quantizing the linear sigma model in the mean field or Hartree
approximation, we have shown how the quantum DCC grows and how it
leads to a squeezed state of the pions.  More generally we have
examined quantum wave functions for the pions from DCC decay,
both of the coherent state and squeezed type.  We have shown that
the classical starting point leads naturally to a quantum
description of the pions in terms of real cartesian (in isospin) pions,
and that that description avoids many of the problem of charge
fluctuations in the final pion state. We have also tried to emphasize
the phenomenological differences among the various quantum descriptions.
We have also discussed how multiple independent domains of DCC,
might manifest themselves.  These are likely to occur in heavy ion
collisions.

Much remains to be done in this growing subject. Better classical and
corresponding quantum models are needed, with a better understanding
of how the classical models connect to particular quantum descriptions.
The notion of solving the difficult strong interaction dynamics
classically and then tying the solution onto a quantum state is a
very promising one and has already shown some success in the
description of nucleon antinucleon annihilation.\cite{Amado1},\cite{Amado2}
However the real test of the DCC concept must come from experiment,
and therefore more phenomenological as well as
purely theoretical work is
need as we approach the time of data.

The work of R. D. Amado is supported in part by the
U. S. National Science Foundation.
The work of I. Kogan was supported U.S. National
 Science Foundation grant NSF PHY90-21984.

\appendix

\section*{Appendix A: Ellipsoidal Isospin Distributions}

All discussions of the neutral fraction proceed from the assumption
of a spherically symmetric a priori distribution in isospin for
the DCC condensate.  For completeness we investigate in this
appendix what happens if we replace this with an ellipsoidal
distribution. We continue to assume that the neutral pion
fraction, $f$, is given by $\cos^2 \theta$,
but now ask for the probability of finding a given $f$ over an
ellipse rather than a sphere.  To do this take ordinary
cartesian coordinates in isospin space.  Then $\cos^2 \theta
=\frac{z^2}{x^2+y^2+z^2}$.  We must take an ellipse that is
symmetric in $x$ and $y$ to preserve zero charge, hence we take
for our ellipse
\begin{equation}
\frac{x^2+y^2}{a^2} +\frac{z^2}{b^2} =1
\end{equation}
The probability of neutral fraction $f$, $P(f)$ is then
\begin{equation}
P(f) = N \int \delta(1-\frac{x^2+y^2}{a^2} -\frac{z^2}{b^2}) \delta(
f-\frac{z^2}{x^2+y^2+z^2}) dx dy dz
\end{equation}
where the normalization is given by
\begin{equation}
1/N =  \int \delta(1-\frac{x^2+y^2}{a^2} -\frac{z^2}{b^2}) dx dy dz
\end{equation}
It is clear that the probability is normalized, $\int P(f) df =1$.
Since $f$ and $P(f)$ are dimensionless, the result for $P(f)$ can
only depend on $\rho = a/b$.  We find
\begin{equation}
P(f)= \frac{\rho}{2 \sqrt{f}} \frac{1}{(a+f(\rho^2-1))^{\frac{3}{2}}}
\end{equation}
This is seen to reduce to the usual spherical answer for $\rho =1$.
It is also easy to see that for $\rho$ near $1$, it is quite close
to the usual answer.  For $\rho$ very large, only very small $f$
dominates (nearly all charged) and for $\rho$ very small, $f$ near
one dominates (nearly all neutral).

\section*{Appendix B: Squeezed and Coherent States}

Squeezed states have been known for a some time
 in quantum optics and measurement theory ( for a review
 of squeezed states see, for example
  \cite{yuen}$^{-}$ \cite{sqrev2}).
 The simplest one-mode squeezed state is parametrized by the
 two parameters $r$ and $\phi$ and can be obtained by  acting
 on the vacuum  with the unitary squeezing operator $S(\xi)$
\begin{equation}
|\xi> = S(\xi)|0> =
 \exp\left[\frac{1}{2}\left(\bar{\xi}a^{2} - \xi (a^{\dagger})^2)
\right)\right]
\end{equation}
where $\xi = r \exp(i\phi)$ is the squeezing parameter.
The mean number of quanta in  the squeezed state is
$\bar{N}= \sinh ^{2} r$.

To see this let us note that using the squeezing operator $S$ one
 can make the Bogoluybov transformation
$b = SaS^{\dagger},~~~
b^{\dagger} = Sa^{\dagger}S^{\dagger}$
 and after some algebra one gets:
\begin{eqnarray}
b &  =  & \cosh r a + \exp(i\phi) \sinh r a^{\dagger},
{}~~~~~~~
b^{\dagger}   = \exp(-i\phi) \sinh r a + \cosh r a^{\dagger}
 \nonumber \\
a &  =  & \cosh r b - \exp(i\phi) \sinh r b^{\dagger},
{}~~~~~~~
a^{\dagger}  = -\exp(-i\phi) \sinh r a + \cosh r a^{\dagger}
\label{abrelation}
\end{eqnarray}
The new operator $b$ is the annihilation operator for the
 squeezed state
\begin{eqnarray}
b|\xi> = b S |0> = S a S^{\dagger} S |0> = S a |0> = 0
\end{eqnarray}
 Then it is easy to see that
\begin{eqnarray}
\bar{N} = <\xi|a^{\dagger}a|\xi>
= \sinh^{2} r <\xi|bb^{+}|\xi> = \sinh^{2} r
\end{eqnarray}

Let us note that
 one   can  rewrite \cite{Hollen}
\begin{equation}
 \exp \left[\frac{1}{2}\left(\bar{\xi}a^{2} - \xi (a^{\dagger})^{2}
\right)\right] =
 \exp \left(\alpha(\xi) a^{\dagger~ 2}\right)
 \exp \left(\beta(\xi) a^{2}\right)
 \exp \left(\gamma(\xi) (a^{\dagger}a+ \frac{1}{2})\right)
\end{equation}
 because $K_{+} = a^{\dagger}a^{\dagger}/2$,
$K_{-} = aa/2$ and $K_{0} = 1/2(a^{+}a + 1/2)$ are the generators
 of the $SU(1,1)$ algebra:
\begin{equation}
[K_{-},K_{+}] = 2K_{0}, ~~~~~
[K_{0}, K_{\pm}] = \pm K_{\pm}
\end{equation}
As a result, the squeezed state can be rewritten as
\begin{eqnarray}
|\xi> = \exp\left(\gamma/2\right)
  \exp \left(\alpha(\xi) a^{\dagger~ 2}\right)|0>
\end{eqnarray}
 To find $\alpha$ and $\gamma$ one has to calculate the
 $\lambda$ dependence of the
 overlap between a squeezed and a coherent state $
<\lambda|\xi> = <0|e^{\lambda a}|\xi>$.
Then
\begin{eqnarray}
\exp\left(\gamma/2\right) = <0|\xi>, ~~~~
\alpha = \frac{1}{2}<0|a^{2} exp \left(\alpha(\xi) a^{\dagger~ 2}\right)|0> =
\frac{1}{2}\exp\left(-\gamma/2\right)
\frac{d^{2}<\lambda|\xi>}{d\lambda^{2}}\vert_{\lambda = 0}
\end{eqnarray}
Using (\ref{abrelation}) twice, and the fact that $b|\xi> = 0$,
 $<0|a^{\dagger} = 0$ and $[\exp(\lambda a), a^{\dagger}] =
\lambda\exp(\lambda a)$ one gets after straightforward calculations
\begin{equation}
\frac{d <\lambda|\xi>}{d\lambda} = - \lambda e^{i\phi}\tanh r
<\lambda|\xi>.
\end{equation}
The solution of this differential equation is
\begin{equation}
 <\lambda|\xi> = \exp\left(-\frac{\lambda^{2}}{2}
e^{i\phi}\tanh r\right)<0|\xi>
\end{equation}
To find $<0|\xi>$ one has to consider the derivative
 $d<0|\xi>/dr $ and again using
 (\ref{abrelation}) twice and   $b|\xi> = 0$,
 $<0|a^{\dagger} = 0$. We find
\begin{eqnarray}
\frac{d}{dr}<0|\xi> = \frac{d}{dr}
<0|\exp \left[\frac{r}{2}\left(\exp(-i\phi)a^{2} - \exp(i\phi)
 (a^{\dagger})^{2}\right)\right]|0> = ~~~~~~~~~~ \nonumber \\
\frac{1}{2} \exp(-i\phi) <0|a^{2}|\xi> =
\frac{1}{2} \exp(+i\phi)\sinh^{2}r <0|b^{\dagger ~2}|\xi> -
 \frac{1}{2}\cosh r \sinh r <0|\xi> = \\
\sinh^{4}r \frac{d}{dr}<0|\xi> +
\frac{1}{2}\cosh r \sinh r (\sinh^{2}r - 1) <0|\xi>
{}~~~~~~~~~~~~~~~~~~~~~~ \nonumber
\end{eqnarray}
and finally
\begin{equation}
\frac{d}{dr}<0|\xi> = -\frac{1}{2} \tanh r <0|\xi>.
\end{equation}
The solution of this differential equation using the boundary
condition, $<0|0>=1$ is
\begin{equation}
<0|\xi> = \frac{1}{\sqrt{\cosh r}}
\end{equation}
The overlap between the squeezed and coherent state is useful
for expressing the squeezed state as a superposition of
coherent states, recalling that the coherent states are
complete.

After a simple calculation we get
\begin{eqnarray}
\alpha = -\tanh r \exp(i\phi),~~~ \exp(\gamma) = \frac{1}{\cosh r}
\end{eqnarray}
Introducing new variable $\alpha = - \tanh r \exp(i\phi)$ one can rewrite
the normalized squeezed state as
\begin{equation}
|\alpha> = (1-|\alpha|^2)^{1/4} \exp \left( \frac{\alpha}{2}
 (a^{\dagger})^2 \right) |0>
\end{equation}
 where $\alpha$
 is a complex number with modulus less than one.  The mean number
of quanta in this state, ${\bar N}$ is given by
 \begin{equation}
<\alpha| a^{\dagger} a |\alpha> = \frac{|\alpha|^2}
{1-|\alpha|^2}
\end{equation}

The squeezed state is a minimum uncertainty state, for position and
momenta defined in terms of the operators $a$:
 $\Delta x\Delta p = 1/2$, where
 $ a(a^{\dagger}) = (x\mp i p)/\sqrt{2}$.
In the case of the coherent state, that minimum uncertainty is achieved
by having each of $\Delta x$ and $\Delta p$ take on their minimum values.
Coherent states  have  minimal quantum noise
 $\Delta x = \Delta p = 1/\sqrt{2}$.
For the squeezed state, only their product need be minimum, and one can
be large while the other is ``squeezed" to keep the product fixed.
 For example for $\phi = 0$ one has $\Delta x = \exp(-r),~~
\Delta p = \exp(r)$.

\end{document}